\documentclass[aps,prd,preprintnumbers,nofootinbib,superscriptaddress]{revtex4}
\usepackage[dvipdfmx]{graphicx}
\usepackage{bm,latexsym,amsmath,amssymb,amsfonts} 
\DeclareMathAlphabet{\mathpzc}{OT1}{pzc}{m}{it}
\usepackage{color}
\usepackage[normalem]{ulem}
\usepackage{hyperref}
\begin{document}
\title{Casimir effect of Lorentz-violating charged Dirac in background magnetic field} 

\author{Ar Rohim}
\email{ar.rohim@ui.ac.id}
\affiliation{Research Center for Quantum Physics, National Research and Innovation Agency (BRIN),\\
South Tangerang 15314, Indonesia}
\affiliation{Departemen Fisika, FMIPA, Universitas Indonesia, Depok 16424, Indonesia}
\author{Apriadi Salim Adam}
\email{apriadi.salim.adam@brin.go.id}
\affiliation{Research Center for Quantum Physics, National Research and Innovation Agency (BRIN),\\
South Tangerang 15314, Indonesia}
\author{Arista Romadani}
\email{arista.romadani@uin-malang.ac.id}
\affiliation{Department of Physics, Faculty of Science and Technology, Universitas Islam Negeri Maulana Malik Ibrahim Malang, Malang  65144, Indonesia}

\begin{abstract}
We study the effect of the Lorentz symmetry breaking on the Casimir energy of charged Dirac in the presence of a uniform magnetic field. We use the boundary condition from the MIT bag model to represent the property of the plates. We investigate two cases of the direction of violation, namely, time-like and space-like vector cases. 
We discuss how the Lorentz violation and the magnetic field affect the structure of the Casimir energy and its pressure. 
We also investigate the weak and strong magnetic field cases with two different limits, heavy and light masses. 

\end{abstract}

\maketitle
%------------------------------------------------
\section{Introduction}
%=======================================================================
 The Casimir effect representing quantum  field effects under macroscopic boundaries was first predicted by H. B. G. Casimir in 1948 \cite{Casimir1948}. He showed that the quantum vacuum fluctuations of the electromagnetic field confined between two parallel plates generate an attractive force. One decade later, in 1958, Sparnaay performed the experimental measurement of the effect, however, with a rough precision \cite{Sparnaay1958}. He found that the attractive force of the plates does not contradict the theoretical prediction.  After his work, the studies showed that the Casimir effect has experimentally confirmed with high precision \cite{Lamoreaux97, Mohideen:1998iz, Roy:1999dx, Bressi:2002fr}. The Casimir effect itself has many applications in nanotechnology \cite{Belluci2009}, and the theoretical discussion was elaborated in connection to several research areas, for example, cosmology \cite{Hassan:2022hcb} and condensed matter physics \cite{Grushin2011, Grushin2021}(see e.g. Refs.~\cite{ Onofrio:2006mq, Bordag:2001qi} for review).

  The studies showed that the Casimir effect also arises not only for the electromagnetic field but also for other fields. The geometry of the plate's surface represented by the form of the boundary conditions
  also determines how the Casimir effect behaves.  To discuss the Casimir effect of the scalar field, one can use the Dirichlet boundary conditions of the vanishing field at the surface of the plates. In such a case, one can also employ Neumann and/or mixed boundary conditions \cite{Ambjorn1983}. 
  However, in the case of the fermion field, one cannot apply such boundaries because the solution for the fermion field is derived from the first-order differential equation. Alternatively, one may use a bag boundary condition that guarantees the vanishing flux at the plate's surface. The well-known form covering this property is the boundary condition from the MIT bag model \cite{Chodos:1974je, Chodos:1974pn} (see Ref.~\cite{Johnson:1975zp} for review). The extension of this boundary that includes the role of the chiral angle has been employed in the literature (see e.g. Refs.~\cite{Rohim:2022mri, Lutken:1983hm}, c.f. Ref.~\cite{Sitenko:2014kza} for the self-adjoint variant).

The Casimir effect phenomenon could be investigated in the system with charged quantum fields under the magnetic field background. With such a system, one can investigate how the charged quantum field couples to the quantum fluctuation \cite{Sitenko:2014kza, Cougo-Pinto:1998jwo, Ostrowski:2005rm, Cougo-Pinto:1998jun, Elizalde:2002kb}. On the other hand, the Casimir effect in the system involving a Lorentz violation has also attracted some attention \cite{Frank:2006ww, Erdas:2013jga, Martin-Ruiz:2016ijc, Cruz:2017kfo, Erdas:2020ilo, Escobar:2020pes, Escobar-Ruiz:2021dxi, Blasone:2018nfy, Cruz:2018thz}.  Within the framework of string theories, the spontaneous Lorentz breaking  may occur through a dynamic of the Lorentz covariant  \cite{Kostelecky:1988zi}. Such a dynamic will generate interactions to gain nonzero expectation values for Lorentz tensors. This is the same analog as in the Higgs mechanism in the context of the standard model. There are several studies where they investigated a system under Lorentz symmetry breaking and the CPT anomaly \cite{Colladay:1996iz, Colladay:1998fq, Kostelecky:2003fs}. Those two phenomena could be possibly measured in the experiment, for instance, the measurements of neutral-meson oscillations \cite{Kostelecky:1994rn, Colladay:1994cj, Colladay:1995qb, Schwingenheuer1995, Gibbons1997, NA31:1990xkc, Kostelecky:1997mh}, the QED test on Penning traps \cite{Schwinberg1981, VanDyck1986, Brown1986, VanDyck1987,  Bluhm:1997ci, Bluhm:1997qb}, and the baryogenesis mechanism \cite{Bertolami:1996cq}. Hence, in this work, we study a system of charged fields involving both Lorentz violation and magnetic field background. In particular, we investigate the Casimir effect of the system under such effects.

 In our setup, the magnetic field is raised in parallel to the normal plate's surface. We investigate two cases of the Lorentz-violating direction, i.e., timelike and space-like directions. For the spacelike case, we restrict ourselves to discussing the violation in the $z$-direction only because the Lorentz violation in the $x$- and $y$-directions do not affect the behavior of the Casimir energy of a Dirac field \cite{Cruz:2017kfo}. In the present study, we employ the boundary condition from the MIT bag model \cite{Chodos:1974je, Chodos:1974pn, Johnson:1975zp}, which is originally used to describe quark confinement. It is natural to show that the presence of the boundary condition in the confinement system leads the allowed perpendicular momentum to the boundary surface to be discrete. To discuss the Casimir effect, we investigate the mode expansion of the field consisting of the linear superposition of the positive- and negative-energy solutions associated with the creation and annihilation operators. We can evaluate the vacuum energy by applying the boundary condition to the mode expansion.  In the present study, we use the Abel-Plana-like summation \cite{Romeo:2000wt} to extract the divergence of the vacuum energy in the presence of boundary conditions. Then, the Casimir energy can be mathematically obtained by taking the difference between the vacuum energy in the presence of the boundary conditions to that in the absence of ones, where both vacuum energies are infinite, but their difference is finite.
  
The rest structure of this paper is organized as follows. In Sec.~\ref{Model}, we describe the model of our system, namely, a Dirac field confined between two parallel plates with a background magnetic field under the Lorentz violation in the quantum field theory framework.  In Sec.~\ref{CasimirEnergy}, we investigate the Casimir energy. 
In this section, we derive the solution for the field inside the confinement area following the procedure used in the literature (see e.g., Refs.~\cite{Bhattacharya:2007vz, Bhattacharya:1999bm, Sitenko:2014kza}). In Sec.~\ref{CasimirPressure}, we discuss the Casimir pressure.  Section~\ref{Summary} is devoted to our summary.  In this paper, we use the natural units so that $c=\hbar=1$.

%=========================================
\section{Model}
\label{Model}
\begin{figure}[tbp]
\centering 
\includegraphics[width=.45\textwidth]{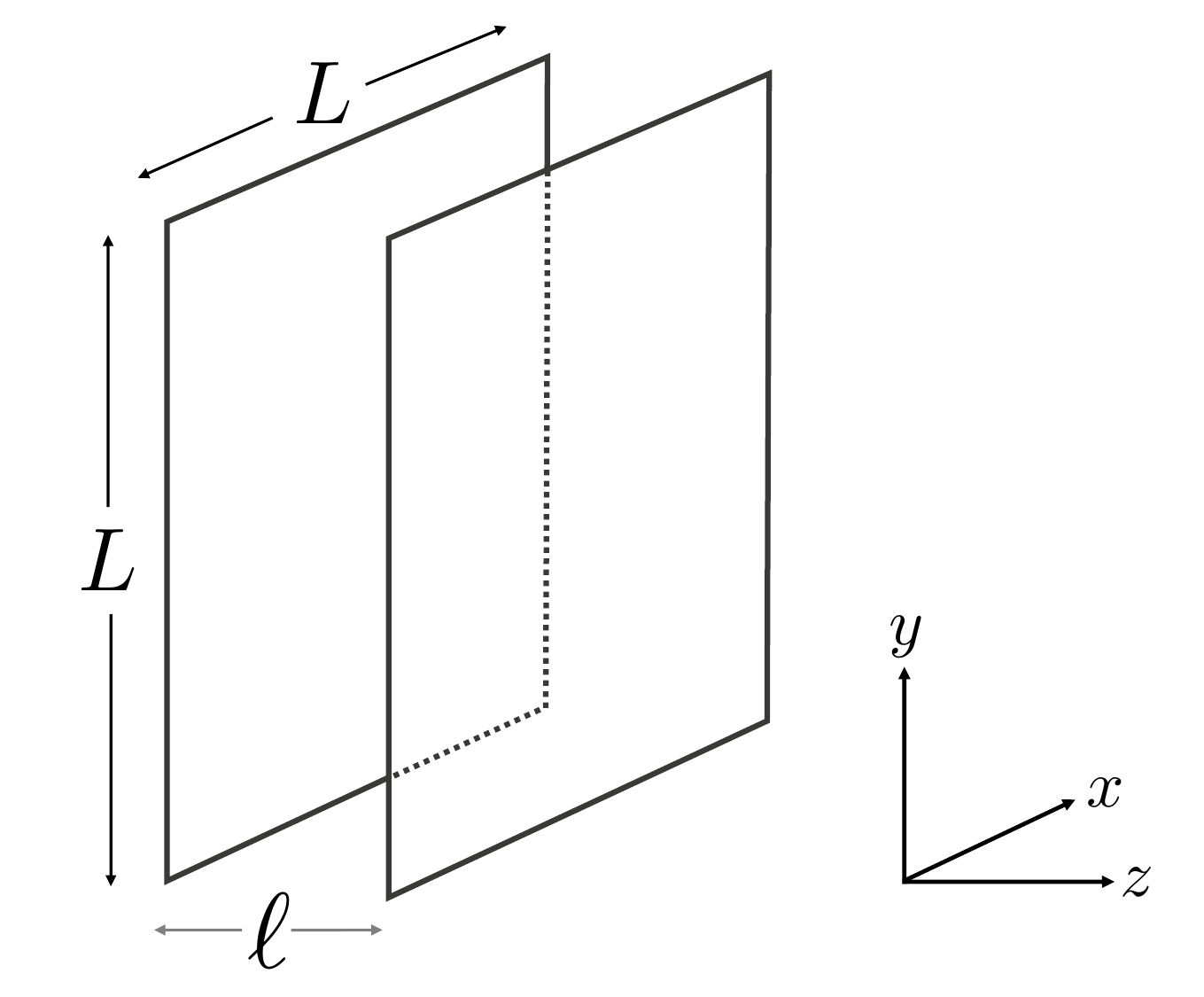}
\caption{\label{Physicalsetup}  Physical setup: Dirac field confined between two parallel plates with the distance $\ell$. The plates with the surface area $L^2$ are placed at $xy$-plane. In this work, we assume the limit $L \rightarrow \infty$ approximately.}
\end{figure}

%=======================================================================
We consider the charged Dirac field confined between two parallel plates placed at $z=0$ and $z=\ell$ in the presence of a uniform magnetic field. The normal surface of the plates is parallel to the $z$-axis (see Fig.~\ref{Physicalsetup}). 
In our model, the Lorentz symmetry is not preserved. The Lagrangian density for such a Dirac field with mass $m$ is given by
\begin{eqnarray}
{\cal L}=\bar\Psi\big[i\gamma^\mu\partial_\mu-e\gamma^\mu A_\mu- m+i\lambda u^\mu u^\nu\gamma_\mu\partial_\nu\big]\Psi,
\label{DiracAction}
\label{LagrangianDensity}
\end{eqnarray}
where $\bar \Psi(\equiv \Psi\gamma^0)$ is the Dirac adjoint, $\lambda$ is the dimensionless parameter with $|\lambda 
|\ll 1$, $A_\mu$ 
is the four vector potential, and $u^\mu$ is an arbitrary constants vector with $u^\mu u_\mu$ can be $1,-1,0$ for time-like, space-like, and light-like, respectively. The Lorentz symmetry breaking is characterized by the last term of Eq.~\eqref{DiracAction}; the parameter $\lambda$ contributes to the violation intensity while the vector $u^\mu$ describes the direction one \cite{Cruz:2018thz}. In the present study, we use the $ 4\times 4$ gamma matrices $\gamma^\mu$ written in the Dirac representation  as follows
\begin{eqnarray}
 \gamma^0=  \begin{pmatrix}
I & 0\\
0 &-I
\end{pmatrix}
~~{\rm and}~~
 \gamma^j=  \begin{pmatrix} 0 & \sigma^j\\
-\sigma^j &0
\end{pmatrix},
\end{eqnarray}
where  $I$ represents the $2\times 2$ identity matrix and $\sigma^j$ is the  $2\times 2$ Pauli matrices.  The gamma matrices satisfy the anti-commutation relation as $\lbrace \gamma^\mu, \gamma^\nu\rbrace=\eta^{\mu\nu}$,  where $\eta^{\mu\nu}(\equiv{\rm diag.}(1,-1,-1,-1))$ is the metric tensor of the Minkowski spacetime. 

The Dirac field $\Psi$ satisfies the modified Dirac equation as follows
\begin{eqnarray}
[i\gamma^\mu\partial_\mu-e\gamma^\mu A_\mu- m+i\lambda u^\mu u^\nu\gamma_\mu\partial_\nu]\Psi=0.
\label{DiracEq}
\end{eqnarray}
The positive-energy solution for the above Dirac equation is given as
\begin{eqnarray}
\Psi^{(+)}(r)=e^{-i\omega t}\psi({\bm r})=e^{-i\omega t}
\begin{pmatrix}
\chi_1\\
\chi_2
\end{pmatrix},
\end{eqnarray}
where $\chi_1$ and $\chi_2$ are the upper and lower two-component spinors, respectively. We use $\omega$ to represent the eigenenergy of the Dirac field.  In our model, the magnetic field is raised in the $z$-direction ${\bm B}=(0,0,B)$, where one can choose the corresponding four-vector potential components as follows 
\begin{eqnarray}
A_0=A_2=A_3=0 ~~~{\rm and}~~~ A_1=-yB, 
\end{eqnarray}
with $B$ as the magnetic field strength. 

The geometry of the plates is described by the boundary condition from the MIT bag model as follows \cite{Chodos:1974je, Chodos:1974pn, Johnson:1975zp}
\begin{eqnarray}
i n_\mu\gamma^\mu\Psi=\Psi,
\label{MITBC}
\end{eqnarray}
where $n_\mu$ is the unit normal inward four-vector perpendicular to the boundary surface. 
The consequence of this boundary is the vanishing flux or normal probability density at the plate surface $n_\mu J^\mu (\equiv n_\mu \bar \Psi \gamma^\mu \Psi)=0$. The idea of this boundary is that the mass of the field is written as a function of its position; inside the confinement area, the mass has a finite value and becomes infinite at the boundary surface. Then, one can suppose that the field outside the confinement area vanishes (see Ref.~\cite{AFG} for the confinement model  of a relativistic particle). While inside the confinement area, the solution for the field is written as the superposition between the left- and right-field components.

%========================================
\section{Casimir energy}
\label{CasimirEnergy}
In this section, we derive the Casimir energy of a Lorentz-violating charge Dirac in a background magnetic field. We study two directions of the Lorentz violation, namely, time-like and space-like vector cases. We derive the solution for the Dirac field inside the confinement area under the boundary condition from the MIT bag model \cite{Chodos:1974je, Chodos:1974pn, Johnson:1975zp}.  
We follow the general procedure given in Refs.~\cite{Bhattacharya:2007vz, Bhattacharya:1999bm, Sitenko:2014kza}. 
Then, we compute the Casimir energy using the Abel-Plana-like summation \cite{Romeo:2000wt} following Refs.~\cite{Bellucci:2009hh, Cruz:2018thz, Rohim:2022mri}. In addition, we also investigate the Casimir energy approximately for the case of weak and strong magnetic fields. 

\subsection{Time-like vector case}
\label{TimeLikeVectorCase}
We consider the positive-energy solution for the timelike vector case with $u^{(t)}=(1,0,0,0)$. In this case, the Dirac equation \eqref{DiracEq} gives two equations as follows
\begin{eqnarray}
&&[(1+\lambda)\omega-m]\chi^{(t)}_1=(-i\sigma^j\partial_j+eyB\sigma^1)\chi^{(t)}_2,\\
&&[(1+\lambda)\omega+m]\chi^{(t)}_2=(-i\sigma^j\partial_j+eyB\sigma^1)\chi^{(t)}_1,
\end{eqnarray}
from which we have the equation for the upper two-component spinor $\chi^{(t)}_1$ as 
\begin{eqnarray}
[(1+\lambda)^2\omega^2-m^2]\chi^{(t)}_1&=&(-i\sigma^j\partial_j+eyB\sigma^1)^2\chi^{(t)}_1\nonumber\\
&=&[-{\bm \nabla}^2+e^2y^2B^2-eB(i2y\partial_1+\sigma^3)]\chi^{(t)}_1. 
\label{DiracEqtimelike}
\end{eqnarray}
In the above equation, we have used the commutation and anti-commutation relations of the Pauli matrices given as   $[\sigma^l,\sigma^m]=2i\epsilon_{lmn}\sigma^n$ and  $\{\sigma^m,\sigma^n\}=2\delta_{mn}I$, respectively,  where $\delta_{mn}$ is a Kronecker delta and $\epsilon_{lmn}$ is a Levi Civita symbol.
To find the solution for $\chi^{(t)}_1$ in Eq.~\eqref{DiracEqtimelike}, one can propose the following form 
\begin{eqnarray}
\chi^{(t)}_1=e^{ik_1 x}e^{ik_3 z} F^{(t)}(y).
\end{eqnarray}
The presence of the Pauli matrix  $\sigma^3$ in Eq.~\eqref{DiracEqtimelike} leads two independent solution for $F^{(t)}(y)$ as follows
\begin{eqnarray}
F^{(t)}_+(y)
=
\begin{pmatrix}
f^{(t)}_+(y)\\
0
\end{pmatrix}
~~~{\rm and}~~~
F^{(t)}_-(y)
=
\begin{pmatrix}
0\\
f^{(t)}_-(y)
\end{pmatrix}
.
\end{eqnarray}
Then, it is convenient to introduce $s=\pm 1$ so that the solution for $f^{(t)}_s(y)$ can be read in a general way as
\begin{eqnarray}
\sigma^3F^{(t)}_s(y)=sF^{(t)}_s(y),
\end{eqnarray}
and introduce a new parameter as
\begin{eqnarray}
\xi^{(+, t)}=\sqrt{eB}\big(y+{k_1 \over eB}\big). 
\end{eqnarray}
Then, Eq.~\eqref{DiracEqtimelike} can be read as  Hermite's equation for arbitrary $s$  as follows
\begin{eqnarray}
\bigg[{d^2\over d\xi^{(t)2}}-\xi^{(t)2}+a^{(t)}_s\bigg]f^{(t)}_s(y)=0,
\label{Hermitetimelike}
\end{eqnarray}
where 
\begin{eqnarray}
a^{(t)}_s={(1+\lambda)^2\omega^2-m^2-k^2_3+eBs\over eB}. 
\end{eqnarray}
We now have the eigenenergies as\footnote{We have used $|eB|$ to avoid imaginary value of $\omega$.}
\begin{eqnarray}
\omega^{(t)}_{n',k_3}=(1+\lambda)^{-1}\sqrt{m^2+k^2_3+|eB|(2n'+1)-|eB|s},
\end{eqnarray}
where we have used $a^{(t)}_s=2n'+1$ with $n'=0,1,2,3,\cdots$. The appropriate solution for $f^{(t)}_s(y)$  with positive value $eB$ that satisfies Hermite's equation \eqref{Hermitetimelike} is given by
\begin{eqnarray}
f^{(t)}_s(y)= \sqrt{{(eB)^{1/2}\over 2^nn'!(\pi)^{1/2}}} e^{-\xi^2/2}H_{n'}(\xi^{(t)}),
\end{eqnarray}
where $f^{(t)}_s(y)$ has been normalized. The solution for $F^{(t)}_s(y)$ is characterized by two conditions, namely, $n'=n$ for $s=+1$ and $n'=n-1$ for $s=-1$. They can be written as follows
\begin{eqnarray}
F^{(t)}_+(y)
=
\begin{pmatrix}
f^{(t)}_{k_1,n}(y)\\
0
\end{pmatrix}
~~~{\rm and}~~~
F^{(t)}_-(y)
=
\begin{pmatrix}
0\\
f^{(t)}_{k_1,n-1}(y)
\end{pmatrix}
.
\end{eqnarray}
We note that the eigenenergy for both values of $s$ gives the same expression as 
\begin{eqnarray}
\omega^{(t)}_{n, k_3}=(1+\lambda)^{-1}\sqrt{m^2+k^2_3+2n|eB|},
\end{eqnarray}
where $n=0,1,2,3,\cdots$ is the Landau level. 
Then, we can finally derive the spatial solution for the right-moving field component as follows
\begin{eqnarray}
\psi^{(+, t)}_{k_1,n,k_3} ({\bm r})&=& {e^{ik_1 x}e^{ik_3 z} \over2\pi\sqrt{2(1+\lambda)\omega^{(t)}_{n, k_3}((1+\lambda) \omega^{(t)}_{n,k_3}+m}) } \nonumber\\
&&\times \left[C_1
  \begin{pmatrix}((1+\lambda)\omega^{(t)}_{n,k_3}+m) f^{(t)}_{k_1, n}(y)\\
  0\\
  k_3f^{(t)}_{k_1, n}(y)\\
  \sqrt{2neB} f^{(t)}_{k_1, n-1}(y)
\end{pmatrix}
+
C_2
  \begin{pmatrix}0\\
  ((1+\lambda)\omega^{(t)}_{n,k_3}+m) f^{(t)}_{k_1, n-1}(y)\\
  \sqrt{2neB} f^{(t)}_{k_1, n}(y)\\
  -k_3f^{(t)}_{k_1, n-1}(y)
\end{pmatrix}
\right],~~{\rm for}~n\geq 1
\end{eqnarray}
\begin{eqnarray}
\psi^{(+, t)}_{k_1,0,k_3} ({\bm r})= {e^{ik_1 x}e^{ik_3 z} \over2\pi\sqrt{2(1+\lambda)\omega^{(t)}_{0, k_3}((1+\lambda) \omega^{(t)}_{0, k_3}+m}) } C_0 f^{(t)}_{k_1, 0}(y)
  \begin{pmatrix}(1+\lambda)\omega^{(t)}_{0, k_3}+m\\
  0\\
  k_3\\
 0
\end{pmatrix},~~{\rm for}~n=0,
\end{eqnarray}
where $C_0, C_1$ and $C_2$ are the complex coefficients and  $f^{(t)}_{k_1, n}(y)$ is given by
\begin{eqnarray}
f^{(t)}_{k_1, n}(y)=\sqrt{{(eB)^{1/2}\over 2^n n!\pi^{1/2}}}
\exp\Bigg[-{eB \over 2} \bigg(y+{k_1\over eB}\bigg)^2\Bigg]H_n\Bigg[\sqrt{eB}\bigg(y+{k_1\over eB}\bigg)\Bigg],
\end{eqnarray}
with $H_n(\xi)$ is the Hermite polynomial. In a similar way, we can obtain the solution for the left-moving field component as follows
\begin{eqnarray}
\psi^{(+, t)}_{k_1,n,-k_3}({\bm r})&=& {e^{ik_1 x}e^{-ik_3 z} \over2\pi\sqrt{2(1+\lambda)\omega^{(t)}_{n, k_3}((1+\lambda) \omega^{(t)}_{n,k_3}+m}) } \nonumber\\ 
&&\times \left[\tilde C_1
  \begin{pmatrix}((1+\lambda)\omega_{nk_3}+m) f^{(t)}_{k_1, n}(y)\\
  0\\
  -k_3f^{(t)}_{k_1, n}(y)\\
  \sqrt{2neB} f^{(t)}_{k_1, n-1}(y)
\end{pmatrix}
+
\tilde C_2
  \begin{pmatrix}0\\
  ((1+\lambda)\omega_{nk_3}+m) f^{(t)}_{k_1, n-1}(y)\\
  \sqrt{2neB} f^{(t)}_{k_1, n}(y)\\
  k_3f^{(t)}_{k_1, n-1}(y)
\end{pmatrix}
\right],~~{\rm for}~n\geq 1
\end{eqnarray}
\begin{eqnarray}
\psi^{(+, t)}_{k_1,0,-k_3} ({\bm r})= {e^{ik_1 x}e^{-ik_3 z} \over2\pi\sqrt{2(1+\lambda)\omega^{(t)}_{0, k_3}((1+\lambda) \omega^{(t)}_{0, k_3}+m}) } \tilde C_0 f^{(t)}_{k_1, 0}(y)
  \begin{pmatrix}
  (1+\lambda)\omega^{(t)}_{0, k_3}+m\\
  0\\
  -k_3\\
 0
\end{pmatrix}
,~~{\rm for}~n=0,
\end{eqnarray}
where $\tilde C_0, \tilde C_1$ and $\tilde C_2$ are the complex coefficients. The total field solution is given by the linear combination between the left- and right-moving field components as follows\footnote{In the case of preserved Lorentz symmetry ($\lambda=0$), the solution is completely the same as that of Ref.~\cite{Sitenko:2014kza}.}
\begin{eqnarray}
\psi^{(+, t)}_{k_1,n, k_3}({\bm r})=\psi^{(+, t)}_{k_1,n,k_{3}}({\bm r})+\psi^{(+, t)}_{k_1,n,-k_{3}}({\bm r}),
\end{eqnarray}
where we use $k_{3 l}$ to represent the allowed momentum in the system, as we will see below.

For arbitrary non-zero complex coefficients, we have the constraint for momentum component in the $z$-direction ($k_3$) in the case of $n\geq 0$ 
as follows 
\begin{eqnarray}
m\ell\sin(k_3\ell)+k_3 \ell \cos (k_3\ell)=0.  
\label{Momentumk3constrainttimelike}
\end{eqnarray}
The detailed derivation is given in Appendix \ref{Detailmomentum}. The solution for Eq.~\eqref{Momentumk3constrainttimelike} is given by $k_{3l}$ with $l=1,2,3,\cdots$, which indicates that the allowed momentum $k_3$ must be discrete.
As a consequence, the energy of the field under the MIT boundary condition must  also be discrete  as follows
\begin{eqnarray}
\omega^{(t)}_{n,l}=(1+\lambda)^{-1}\sqrt{m^2+k^2_{3l}+2n|eB|}.
\end{eqnarray}
These properties not only hold for positive-energy solutions but also for the negative-energy counterpart. One can see that the magnetic field and parameter $\lambda$ do not affect the structure of the momentum constraint. In this context, the former is similar to that in the absence of the magnetic field \cite{Cruz:2018thz} while the latter is similar to that of the preserved Lorentz symmetry. 

We now write down a mode expansion of the Dirac field in the time-like vector case under the boundary condition from the MIT bag model as
\begin{eqnarray}
\Psi^{(t)}_{k_1,n,l}(r)= \sum^\infty_{n=0}\sum^\infty_{l=1}\int^\infty_{-\infty}d k_1
\big[\hat a_{k_1,n,l} \Psi^{(+,t)}_{k_1,n,l}(r)+ \hat b^\dagger_{k_1,n,l} \Psi^{(-,t)}_{k_1,n,l}(r) \big],
\label{ModeTimeLike}
\end{eqnarray}
where $\Psi^{(\pm,t)}_{k_1,n,l}(r)$ are the positive (+) and negative (-) energy solutions. See Appendix~\ref{NETimelikevectorcase} for the detailed expression of the negative-energy solution. The annihilation and creation operators in Eq.~\eqref{ModeTimeLike} satisfy the following anti-commutation relations
\begin{eqnarray}
\{\hat a_{k_1,n,l},\hat a^\dagger_{k'_1,n',l'}\}=\{\hat b_{k_1,n,l},\hat b^\dagger_{k'_1,n',l'}\}=\delta_{nn'}\delta_{ll'}\delta(k_1-k'_1),
\end{eqnarray}
and the other anticommutation relations vanish. The Dirac field satisfies  orthonormality conditions as follows
\begin{eqnarray}
\int d{\bm x}_{\perp}\int^\ell_0 dz \psi^{(j,t)\dagger}_{k_1,n, l}({\bm r})\psi^{(j',t)}_{k'_1,n', l'}({\bm r})=\delta_{jj'}\delta_{nn'}\delta_{l l'}\delta(k_1-k'_1),~~~~j,j'=0,1,2~,
\label{OrthonormalityTimeLike}
\end{eqnarray}
by which we can obtain the relations of the complex coefficients of the field. 
We use ${\bm x}_{\perp}\equiv (x,y)$ to represent the sub-spatial coordinate parallel to the normal plates' surface. 
From the above Lagrangian density \eqref{LagrangianDensity}, one can obtain the Hamiltonian density in the time-like vector case as follows
\begin{eqnarray}
{\cal H}^{(t)}=-\bar\Psi^{(t)}\big[i\gamma^j\partial_j-e\gamma^\mu A_\mu- m\big]\Psi^{(t)}=i(1+\lambda)\Psi^{(t)\dagger}\partial_0\Psi^{(t)}.
\end{eqnarray}
Then we are now ready to evaluate the vacuum energy as follows 
\begin{eqnarray}
E^{(t)}_{\rm Vac.}=\int_\Omega d^3 {\bm x}{\cal E}^{(t)}_{\rm Vac.}=\int_\Omega d^3 {\bm x} \langle 0|{\cal H}^{(t)}|0\rangle = -{|eB|L^2\over \pi } \sum_{n=0}^\infty \sum_{l=1}^\infty i_n\sqrt{m^2+\bigg({k'_{3l}\over \ell}\bigg)^2+2n|eB|},
\end{eqnarray}
where ${\cal E}_{\rm Vac.}$ is the vacuum energy density, $i_n=1-{1\over 2}\delta_{n0}$, $k'_{3l}\equiv k_{3l}\ell$, and $\Omega$ is the volume of the confinement area. One can derive the Casimir energy by subtracting the vacuum energy in the presence of the boundary condition from the absence of one.  
We note that the roles of $\lambda$ do not appear in the vacuum energy for the time-like vector case. In other words, the Casimir energy also does not depend on $\lambda$. In the next subsection, we will show that the above result can be recovered in the case of the preserved Lorentz symmetry. Therefore, it is not necessary to evaluate further the Casimir energy in this subsection.

%=========================================================

\subsection{Space-like vector case} 
\label{SpaceLikeVectorCase}

In this subsection, we investigate the Casimir energy for the space-like vector case in the $z$-direction. 
We start the discussion by deriving the solution for the space-like vector case with $u^{(z)}=(0,0,0,1)$.  In this case, the Dirac equation \eqref{DiracEq} gives two equations as follows
\begin{eqnarray}
&&(\omega-m)\chi^{(z)}_1=(-i\sigma^j\partial_j+eyB\sigma^1+i\lambda\sigma^3\partial_3)\chi^{(z)}_2,
\label{EqDirac1Spacelike}\\
&&(\omega+m)\chi^{(z)}_2=(-i\sigma^j\partial_j+eyB\sigma^1+i\lambda\sigma^3\partial_3)\chi^{(z)}_1.
\label{EqDirac2Spacelike}
\end{eqnarray}
Multiplying both sides of Eq.~\eqref{EqDirac1Spacelike} by $(\omega+m)$ and using Eq.~\eqref{EqDirac2Spacelike}, we have the equation for the upper two-component spinor $\chi^{(z)}_1$ as follows
\begin{eqnarray}
(\omega^2-m^2)\chi^{(z)}_1&=&(-i\sigma^j\partial_j+eyB\sigma^1+i\lambda\sigma^3\partial_3)^2\chi^{(z)}_1\nonumber\\
&=& [-{\bm \nabla}^2+e^2y^2B^2-eB(2iy\partial_1+\sigma^3)+2\lambda \partial^2_3-\lambda^2\partial^2_3]\chi^{(z)}_1.
\label{EqChi1z}
\end{eqnarray}
One can propose the solution $\chi^{(z)}_1$ as follows 
\begin{eqnarray}
\chi^{(z)}_1=e^{ik_1 x}e^{ik_3 z}f^{(z)}(y).
\label{SolSpacelike}
\end{eqnarray}
Along the same procedure used in the previous subsection, substituting back Eq.~\eqref{SolSpacelike} into Eq.~\eqref{EqChi1z} brings us to  Hermite's equation in which we have the eigen energies given as
\begin{eqnarray}
\omega^{(z)}_{n, k_3}=\sqrt{m^2+(1-\lambda)^2k^2_3+2 n |eB|}.
\label{Energyz}
\end{eqnarray}
We find that the solution of the Dirac field confined between two parallel plates in the space-like vector case of $z$-direction for the right-moving field with positive value $eB$ is given as follows
\begin{eqnarray}
\psi^{(z)}_{k_1,n,k_3} ({\bm r})={e^{ik_1 x}e^{ik_3 z} \over2\pi\sqrt{2\omega^{(z)}_{n, k_3}(\omega^{(z)}_{n, k_3}+m}) } \left[C_1
  \begin{pmatrix}(\omega_{n, k_3}+m) F^{(z)}_{k_1, n}(y)\\
  0\\
(1-\lambda)  k_3F^{(z)}_{k_1, n}(y)\\
  \sqrt{2neB} F^{(z)}_{k_1, n-1}(y)
\end{pmatrix}
+
C_2
  \begin{pmatrix}0\\
  (\omega_{nk_3}+m) F^{(z)}_{k_1, n-1}(y)\\
  \sqrt{2neB} F^{(z)}_{k_1, n}(y)\\
  -(1-\lambda)k_3F^{(z)}_{k_1, n-1}(y)
\end{pmatrix}
\right],~~{\rm for}~n\geq 1\nonumber\\
\end{eqnarray}
\begin{eqnarray}
\psi^{(z)}_{k_1,0, k_3} ({\bm r})= {e^{ik_1 x}e^{ik_3 z} \over2\pi\sqrt{2\omega^{(z)}_{0, k_3}(\omega^{(z)}_{0, k_3}+m}) }  C_0 F^{(z)}_{k_1 0}(y)
  \begin{pmatrix}\omega^{(z)}_{0, k_3}+m\\
  0\\
(1-\lambda)  k_3\\
 0
\end{pmatrix}
,~{\rm for}~n=0,
\end{eqnarray}
where
\begin{eqnarray}
F^{(z)}_{k_1, n}(y)=\sqrt{{(eB)^{1/2}\over 2^n n!\pi^{1/2}}}
\exp\Bigg[-{eB \over 2} \bigg(y+{k_1\over eB}\bigg)^2\Bigg]H_n\Bigg[\sqrt{eB}\bigg(y+{k_1\over eB}\bigg)\Bigg],
\end{eqnarray}
with the Hermite polynomial $H_{n}(y)$. In a similar way, we can obtain the solution  for the left-moving field as follows
\begin{eqnarray}
\psi^{(+,z)}_{k_1,n,-k_3}({\bm r})= {e^{ik_1 x}e^{-ik_3 z} \over2\pi\sqrt{2\omega^{(z)}_{n, k_3}(\omega^{(z)}_{n, k_3}+m}) }  \left[\tilde C_1
  \begin{pmatrix}(\omega^{(z)}_{n, k_3}+m) F^{(z)}_{k_1, n}(y)\\
  0\\
  -(1-\lambda)k_3F^{(z)}_{k_1, n}(y)\\
  \sqrt{2neB} F^{(z)}_{k_1, n-1}(y)
\end{pmatrix}
+
\tilde C_2
  \begin{pmatrix}0\\
  (\omega^{(z)}_{n, k_3}+m) F^{(z)}_{k_1, n-1}(y)\\
  \sqrt{2neB} F^{(z)}_{k_1, n}(y)\\
  (1-\lambda)k_3F^{(z)}_{k_1, n-1}(y)
\end{pmatrix}
\right],~{\rm for}~n\geq 1\nonumber\\
\end{eqnarray}
\begin{eqnarray}
\psi^{(+,z)}_{k_1,0,-k_3} ({\bm r})= {e^{ik_1 x}e^{-ik_3 z} \over2\pi\sqrt{2\omega^{(z)}_{0, k_3}(\omega^{(z)}_{0, k_3}+m}) } \tilde C_0 F^{(z)}_{k_1, 0}(y)
  \begin{pmatrix}\omega^{(z)}_{0,k_3}+m\\
  0\\
  -(1-\lambda)k_3\\
 0
\end{pmatrix}
,~{\rm for}~n=0,
\end{eqnarray}
where the eigen energies $  \omega^{(z)}_{n,k_3}$ are given by Eq.~\eqref{Energyz} (see Appendix \ref{Detailmomentum} for the detailed derivation). The complex coefficients in the above Dirac field can be determined by similar orthonormality conditions given in Eq.~\eqref{OrthonormalityTimeLike}.

We next write the total spatial solution for the Dirac field inside the confinement area as follows
\begin{eqnarray}
\psi^{(+,z)}_{k_1,n,k_3}({\bm r})=\psi^{(+,z)}_{k_1,n,k_{3}}({\bm r})+\psi^{(+,z)}_{k_1,n,-k_{3}}({\bm r}).
\end{eqnarray}
For non-zero complex coefficients $C_1, C_2,\tilde C_1,\tilde C_2$, we have the constraint of the momentum $k_3$ as follows
\begin{eqnarray}
m\ell\sin(k_3\ell)+(1-\lambda)k_3 \ell \cos (k_3\ell)=0, 
\label{MomentumConstraintz}
\end{eqnarray}
for arbitrary Landau level $n$. One can see that the parameter $\lambda$  affects the constraint while the magnetic field does not. The allowed momentum that satisfies the constraint \eqref{MomentumConstraintz} is $k_{3l}$ with $l=0,1,2,3,\cdots$. The discretized eigenenergies of the field under the MIT boundary can be written  as follows
\begin{eqnarray}
\omega^{(z)}_{n,l}=\sqrt{m^2+(1-\lambda)^2 k^2_{3l}+2n|eB|}.
\label{Energyz2}
\end{eqnarray}

Below we will compute the Casimir energy of charged Dirac field under the presence of the MIT boundary. For this purpose, we write down the Hamiltonian density for the space-like vector case as follows,
\begin{eqnarray}
{\cal H}^{(z)}=-\bar\Psi^{(z)}\big[i\gamma^j\partial_j-e\gamma^\mu A_\mu- m\big]\Psi^{(z)}=i\Psi^{(z)\dagger}\partial_0\Psi^{(z)}.
\end{eqnarray}
The vacuum energy reads
\begin{eqnarray}
E_{\rm Vac.}=-{|eB| L^2\over \pi}
\sum_{n=0}^\infty \sum_{l=1}^\infty i_n \sqrt{m^2+(1-\lambda)^2\bigg({k'_{3l}\over \ell}\bigg)^2+2n|eB|},
\label{Vacuumzdirection}
\end{eqnarray}
where we have used the eigenenergies given in Eq.~\eqref{Energyz2} and $k'_{3\ell}(\equiv k_{3l}\ell)$. From the above vacuum energy, one can see that its value is divergent. To solve the issue, we employ the Abel-Plana-like summation as follows \cite{Romeo:2000wt}
\begin{eqnarray}
\sum_{l=1}^\infty {\pi f_n(k'_{3l})\over \bigg(1-{\sin(2k'_{3l})\over 2k'_{3l}}\bigg)}=-{\pi b mf_n(0)\over 2 (b m+1)}+\int_0^\infty dz f_n(z) - i\int_0^\infty dt {f_n (it)-f_n(-it)\over {t+b m\over t-b m}e^{2t}+1}.
\label{denumerator}
\end{eqnarray}
From the momentum constraint in the space-like vector case \eqref{MomentumConstraintz}, the denominator of the left-hand side Eq.~\eqref{denumerator} can be rewritten in the following form
\begin{eqnarray}
1-{\sin(2k'_{3l})\over 2k'_{3l}} =  1 +{b m\over k'^2_{3l}+(bm)^2},
\end{eqnarray}
where 
\begin{eqnarray}
b=\ell (1-\lambda)^{-1}. 
\end{eqnarray}
Then, after applying the Abel-Plana-like summation to the vacuum energy, Eq.~\eqref{Vacuumzdirection} becomes 
\begin{eqnarray}
E_{\rm Vac.}=-{|eB|L^2\over \pi^2 b}
\sum_{n=0}^\infty i_n \bigg[-{\pi b m f_n(0)\over 2 (b m +1)}+\int_0^\infty dq f_n(q) - i\int_0^\infty dt {f_n (it)-f_n(-it)\over {t+b m\over t-b m}e^{2t}+1}\bigg],
\label{ECasAbelPlana}
\end{eqnarray}
where the function $f_n(q)$ is defined as 
\begin{eqnarray}
f_n(q)= \sqrt{m^2b^2+q^2+2n|eB| b^2} \bigg(1 +{b m\over q^2+(bm)^2}\bigg).
\label{fq}
\end{eqnarray}
Next, one can decompose the first and second terms in the vacuum energy \eqref{ECasAbelPlana} into two parts: (i) in the absence of the boundary conditions of two plates and  (ii) in the presence of one plate. The latter part is irrelevant to our discussion because it does not contribute to the force. Then, the last term of Eq.~\eqref{ECasAbelPlana} can be understood as the Casimir energy
\begin{eqnarray}
E_{\rm Cas.}={i |eB|L^2\over \pi^2 b}
\sum_{n=0}^\infty i_n \int_0^\infty dt {f_n (it)-f_n(-it)\over {t+b m\over t-b m}e^{2t}+1}.
\end{eqnarray}
Using Eq.~\eqref{fq} and introducing variable of $t=bu$, the Casimir energy reads
\begin{eqnarray}
   E_{\rm Cas.}=  -\frac{2 |e B| L^2 }{\pi^2 } \sum_{n = 0}^{\infty} i_n \int_{0}^{\infty} d u
  \sqrt{u^2 - M_n^2 } \left( \frac{b \left( u  - m \right)   - m /   (m +
  u)}{(u + m) e^{2 b u} + u - m} \right),
  \label{ECas1}
\end{eqnarray}
where 
\begin{eqnarray}
\begin{array}{lll}
  M_n &=&\sqrt{m^2 + 2 n |e B|}. 
\end{array}
\end{eqnarray}
The range of integration of Eq.~\eqref{ECas1} can be split into two intervals, i.e., $[0,M_n]$  and $[M_n,\infty]$. The integration result of the first interval vanishes while the second one remains. To further proceed with the Casimir energy,  
we next rewrite the following quantity as
\begin{eqnarray}
  \frac{b {(u  - m) }  - m / (m + u)}{(u + m) e^{2 b u} + u - m} & = & -
  \frac{1}{2} \frac{d}{d u} \ln \left( 1 + \frac{u - m}{u + m} e^{- 2 b u}
  \right),
\end{eqnarray}
which leads the Casimir energy to
\begin{eqnarray}
  E_{\rm Cas.} = \frac{|e B| L^2 }{ \pi^2 b} \sum_{n = 0}^{\infty} i_n \int_0^{\infty} d y
  \sqrt{y^2 + 2 y b M_n} \frac{d}{d y} \ln \left( 1 + \frac{y + b (M_n - m)}{y + b
  (M_n + m)} e^{- 2 (y + b M_n) } \right),
  \label{EC}
\end{eqnarray}
where we have introduced a new variable as
\begin{eqnarray}
  y  = b u - b M_n.
\end{eqnarray}
Performing integration by part for Eq.~\eqref{EC}, we finally find the simpler form of the Casimir energy as follows
\begin{eqnarray}
E_{\rm Cas.}=-{|eB|
L^2\over \pi^2 b}\sum^\infty_{n=0} i_n \int^\infty_0 dy (y+bM_n)(y^2+2byM_n)^{-1/2}\ln \left( 1 + \frac{y + b (M_n - m)}{y + b (M_n + m)} e^{- 2 (y + b M_n) }
\right).
\label{ECas}
\end{eqnarray}

\begin{figure}[tbp]
\centering 
\includegraphics[width=.49\textwidth]{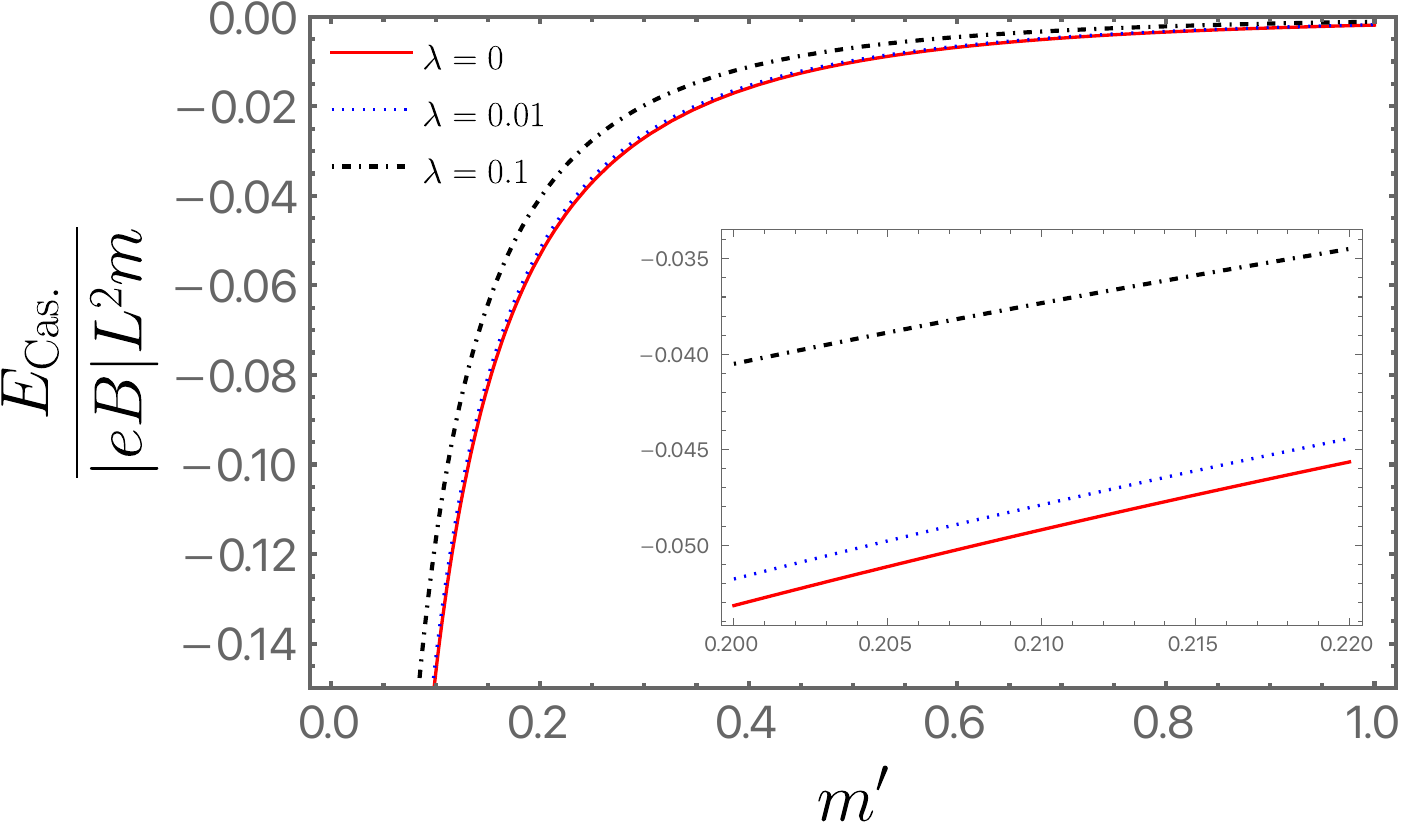}
\hfill
\includegraphics[width=.49\textwidth]{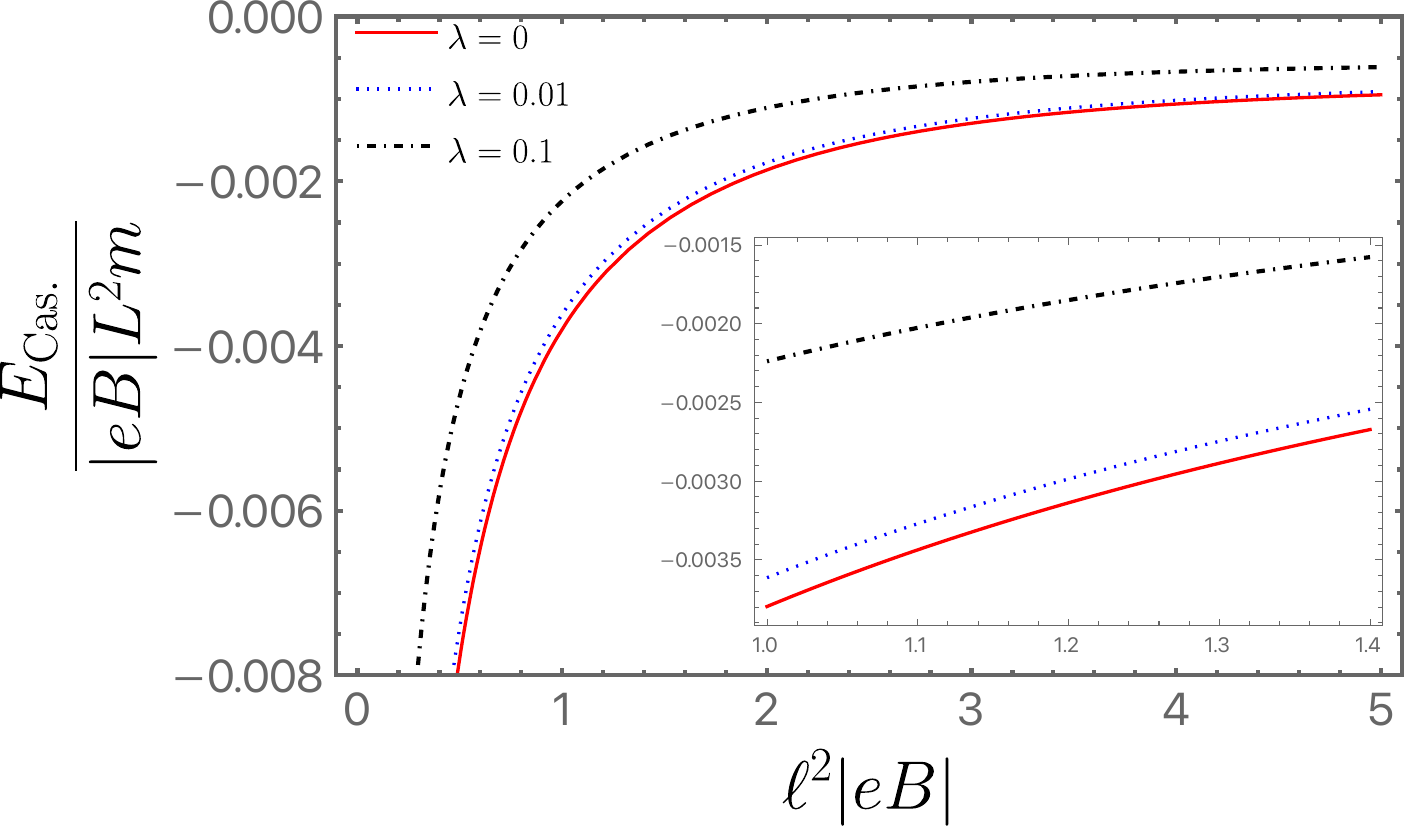}
\caption{\label{EnCas1} The left panel plots  the scaled Casimir energy  of the space-like vector case as a function of the dimensionless parameter $m'$ for fixed $\ell^2|eB|=2$.  While the right panel is the scaled Casimir energy  of the space-like vector case as a function of the dimensionless parameter $\ell^2 |eB|$ for fixed $m'=1$. In both panels, we use three values of the parameter $\lambda=0,0.01,0.1$. }
\end{figure}

\begin{figure}[tbp]
\centering 
\includegraphics[width=.49\textwidth]{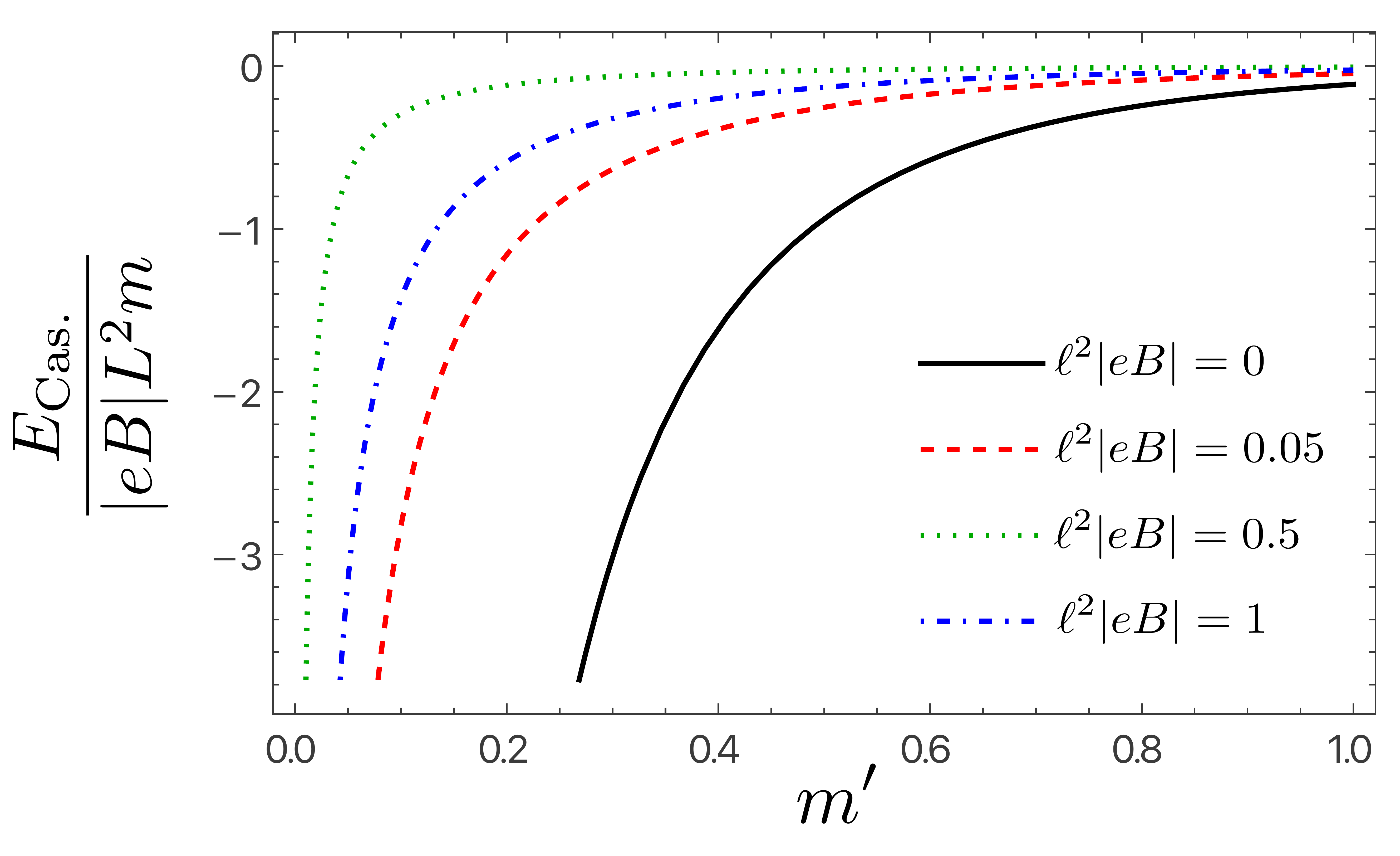}
\caption{\label{EnCas3} The scaled Casimir energy  of the space-like vector case as a function of the dimensionless parameter $m'(\equiv m \ell)$ for various values of parameter $\ell^2 |eB|=0,0.05,0.5,1$ and fixed parameter $\lambda=0.1$.}
\end{figure}

We next numerically evaluate the expression of the Casimir energy given in Eq.~\eqref{ECas}. The left panel of Fig.~\ref{EnCas1} depicts the scaled Casimir energy as a function of the dimensionless parameter $m'(\equiv m\ell)$ for various values of the parameter $\lambda=0,0.01,0.1$ with a fixed parameter $\ell^2|eB|=2$.   From this figure, we find that the scaled Casimir energy converges to zero as the parameter $m'$ becomes larger. The right panel of figure~\ref{EnCas1} depicts the scaled Casimir energy as a function of the dimensionless parameter $\ell^2|eB|$ for a fixed parameter $m'=1$. From this figure, one can see that the scaled Casimir energy also converges to zero as the parameter $\ell^2 |eB|$ increases. Both panels of Fig.~\ref{EnCas1} show that the parameter $\lambda$ increases, the Casimir energy will increase and vice versa, as previously shown by Ref.~\cite{Cruz:2018thz} for the absence of the magnetic field.  
Figure \ref{EnCas3} plots the scaled Casimir energy as a function of the dimensionless parameter $\ell^2 |eB|$ for various values of parameter $\lambda=0,0.01,0.1$ with a fixed parameter $m'=1$. One can see that the increasing $\ell^2 |e B|$ leads to the converging of the Casimir energy to zero.


In the rest of this part, we investigate the approximate cases of the Casimir energy.  In the  case of the weak magnetic field $B\to 0$, the above Casimir energy \eqref{ECas} for an arbitrary $m'(\equiv m\ell)$  reduces to 
\begin{eqnarray}
E_{\rm Cas.}\simeq -{L^2\over \pi^2 b^3}\int^\infty_{bm} dx x^2 \int^\infty_0 dv (v+1){1\over \sqrt{v(v+2)}}\ln \left( 1+{x(v+1)-bm\over x(v+1)+bm } e^{-2x(v+1)}\right).
\label{ECasB0}
\end{eqnarray}
To obtain the above expression, we have used the replacement of summation with integration, $v=y/(b M_n)$, and $x=bM_n$. Taking the case of light mass $m'\ll1$ for  Eq.~\eqref{ECasB0}, we recover the earlier  result by Ref.~\cite{Cruz:2018thz} as follows
\begin{eqnarray}
E_{\rm Cas.}\simeq-{7\pi^2 (1-\lambda)^3 L^2\over 2880  \ell^3}\left[1-{120 m'\over 7\pi^2(1-\lambda)}\right],
\label{ECasB0mll1}
\end{eqnarray}
where we have expanded the integrand up to the order of $\mathcal{O}(m')$ and omitted the higher ones. The first term corresponds to the Casimir energy in the massless case with the effect of the Lorentz violation while the second term corresponds to the correction part.  In the case of the preserved Lorentz symmetry, $\lambda=0$, we recover the well-known Casimir energy of the massless fermion derived by Johnson \cite{Johnson:1975zp}.

To obtain the approximated result of Eq.~\eqref{ECasB0mll1}, one can also start from the general Casimir energy \eqref{ECas} and take its light mass  case $m'\ll 1$ for the arbitrary magnetic field as
\begin{eqnarray}
    E_{\rm Cas.}\simeq -{|eB|L^2\over \pi^2b}\sum_{n=0}^\infty i_n\int^\infty _0dy \left[{(y+b\sqrt{2 n e B})\ln{(1+e^{-2(y+b\sqrt{2 n e B})})}\over \sqrt{y^2+2y b\sqrt{2n e B}}}-{2b me^{-2(y+b\sqrt{2 n e B})}\over \sqrt{y^2+2y b\sqrt{2n e B}}(1+e^{-2(y+b\sqrt{2 n e B})})}\right].\nonumber\\
    \label{ECasmll1}
\end{eqnarray}
Then, taking the limit of the weak magnetic field, the above expression reduces to Eq.~\eqref{ECasB0mll1}.

In the case of heavy mass  $m'\gg 1$, we find that the Casimir energy approximately reduces to
\begin{eqnarray}
E_{\rm Cas.} \simeq - {|e B|L^2(1-\lambda)^{3/2}\over 16 \pi^{3/2}\ell \sqrt{m'}}\sum_{n=0}^ \infty i_ne^{-2\sqrt{  m'^2+2 n B'}\over (1-\lambda)},
\label{ECasmpgg1}
\end{eqnarray}
where we have expanded the integrand of Eq.~\eqref{ECas} up to the order of $\mathcal{O}(1/m')$ and omitted the higher ones. In the case of weak magnetic field $B\to 0$, the above Casimir energy \eqref{ECasmpgg1} reads
\begin{eqnarray}
    E_{\rm Cas.}\simeq - \frac{L^2 (1 - \lambda)^{5 / 2} \sqrt{m'}}{32 \pi^{3 / 2} \ell^3 } e^{-
\frac{{2 m'} }{(1 - \lambda)} }.
\label{ECasgg1}
\end{eqnarray}
We can see that, in the case of heavy mass, the Casimir energy goes to zero as the increase of mass. 

We next investigate the Casimir energy in the case of the strong magnetic field $\ell^2 eB\gg 1$. In this case,  together with light mass $m'\ll 1$, the Casimir energy in Eq.~\eqref{ECas} approximately reduces to 
\begin{eqnarray}
E_{\rm Cas.}\simeq -{|eB|L^2 (1-\lambda) \over 48 \ell}.
\label{ECasll1}
\end{eqnarray}
Meanwhile for the case of strong magnetic field $\ell^2 |eB|\gg 1$ and taking the limit of heavy mass $m'\gg 1$, the Casimir energy reads
\begin{eqnarray}
E_{\rm Cas.}\simeq-{|eB|L^2 (1-\lambda)^{3/2} \over 32 \pi^{3/2} \ell \sqrt{m'}}e^{-2m'\over (1-\lambda)}.
\label{ECasgg1}
\end{eqnarray}
From the above expression, we note that the Casimir energy converges to zero as the increase of parameter $m'$.

%=============================================================================
\section{Casimir pressure}
\label{CasimirPressure}
%=============================================================================

\begin{figure}[tbp]
\centering 
\includegraphics[width=.48\textwidth]{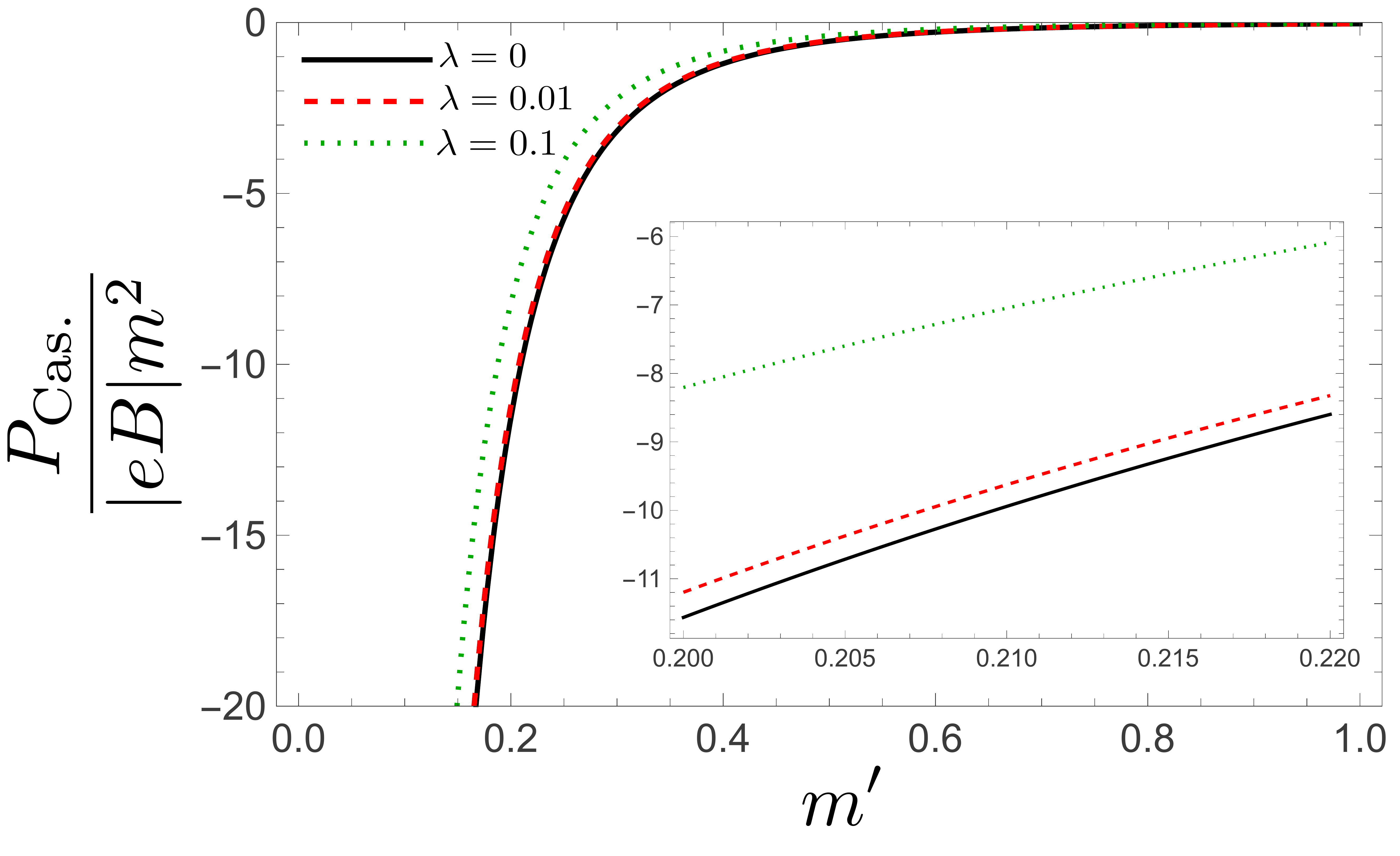}
\hfill
\includegraphics[width=.50\textwidth]{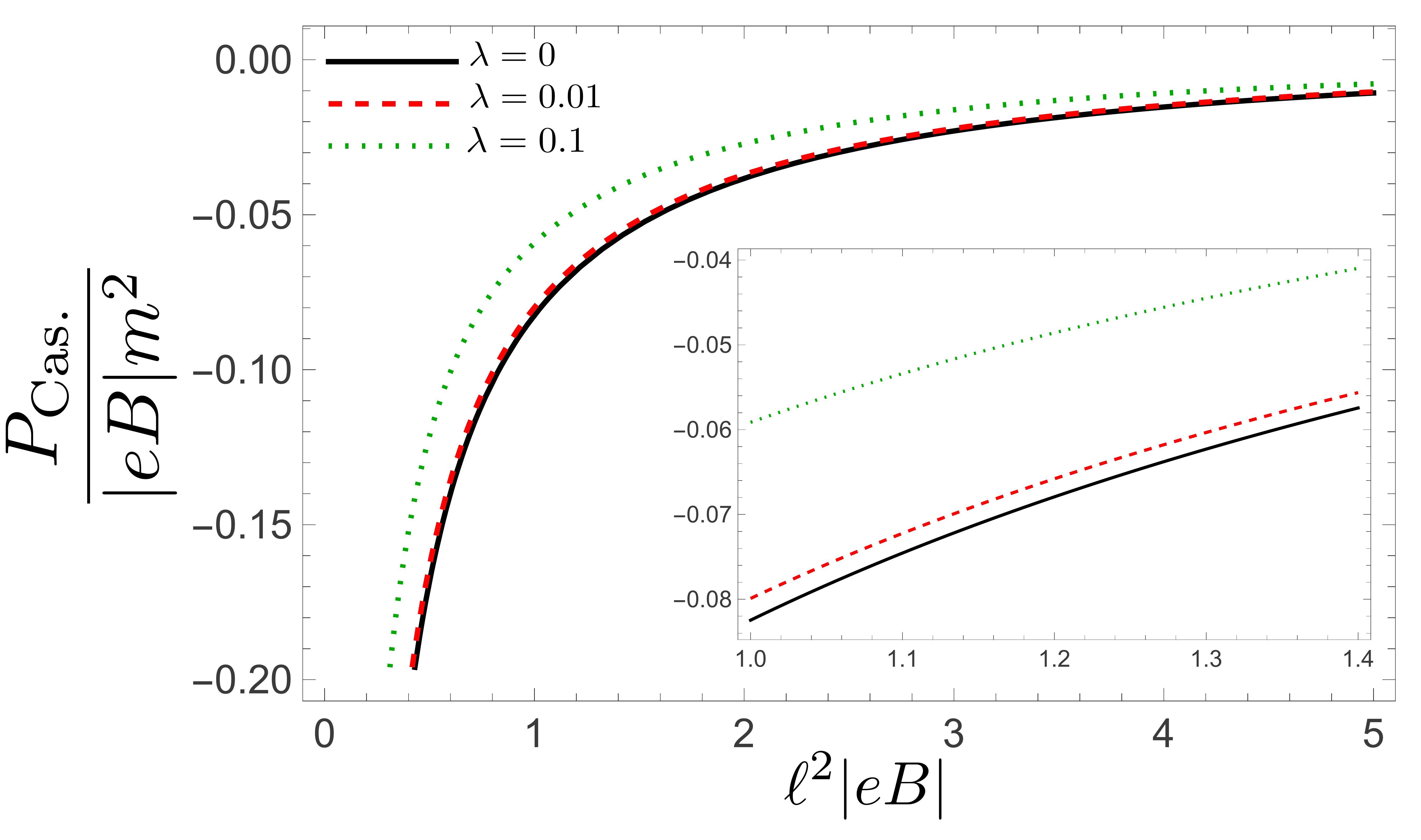}
\caption{\label{PCas1} The left panel plots the scaled Casimir pressure of the spacelike vector case as a function of the dimensionless parameter $m'$ for fixed $\ell^2|eB|=2$.  While the right panel is the scaled Casimir pressure of the spacelike vector case as a function of the dimensionless parameter $\ell^2 |eB|$ for fixed $m'=1$. In both panels, we use the various values of the parameter $\lambda=0,0.01,0.1$. }
\end{figure}

\begin{figure}[tbp]
\centering 
\includegraphics[width=.49\textwidth]{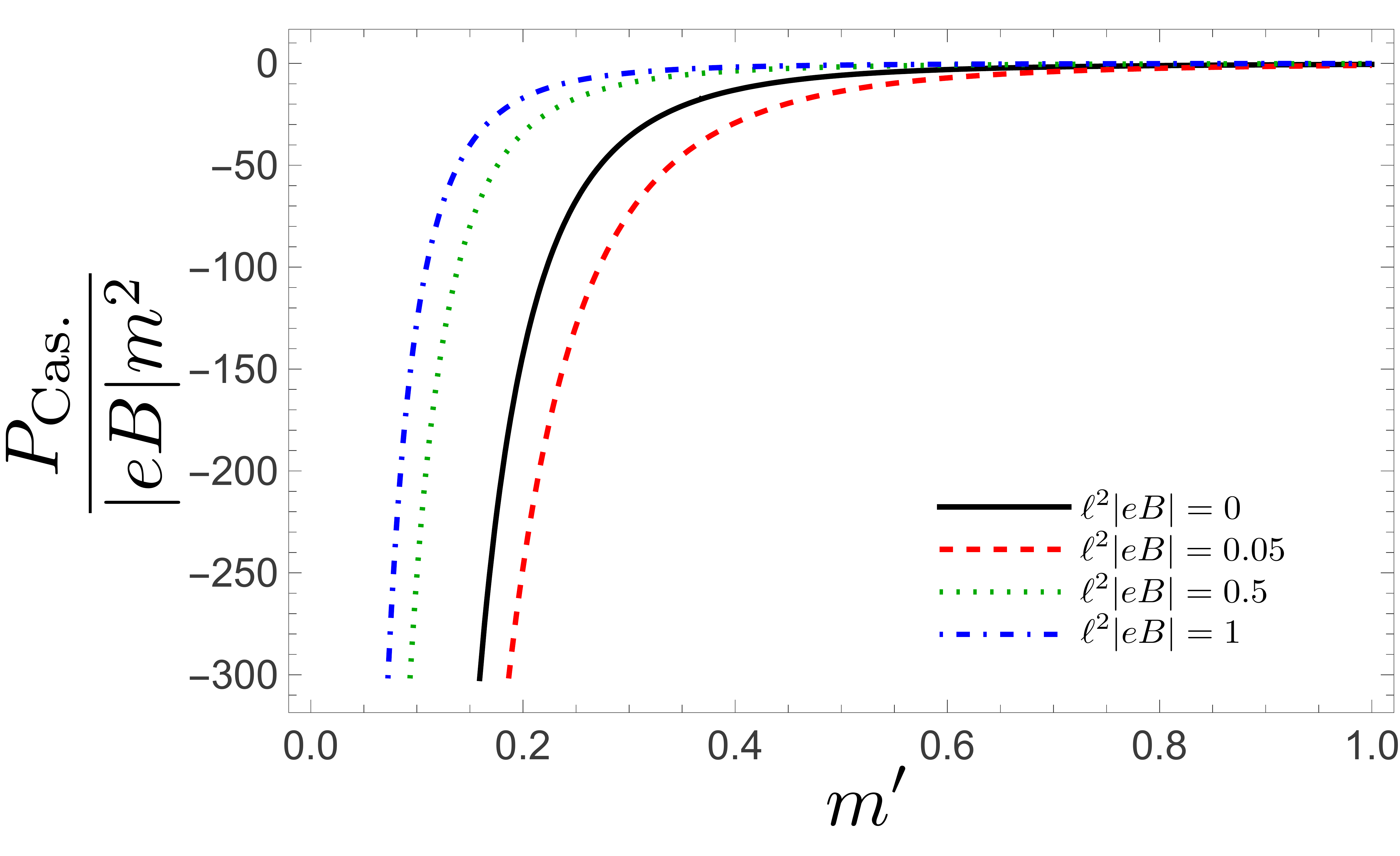}
\caption{\label{PCas3} The scaled Casimir pressure of the spacelike vector case as a function of the dimensionless parameter $m'(\equiv m \ell)$ for various values of parameter $\ell^2 |eB|=0,0.05,0.5,1$ and fixed $\lambda=0.1$.   }
\end{figure}

In this section, we investigate the Casimir pressure for the spacelike vector case. It can be obtained from the Casimir energy \eqref{ECas}  by taking the derivative with respect to the plate's distance as
\begin{eqnarray} 
  P_{\text{Cas} .} & =& -{1\over L^2}{\partial E_{\rm Cas.} \over \partial \ell}\nonumber\\
  &=&  - \sum_{n = 0}^{\infty} i_n \int_0^{\infty} d y
  \frac{1}{(1 - \lambda) b^2 \pi^2 (y (2 b M_n + y))^{3 / 2} }\nonumber\\
  &  & ~~~~~~~~~~~~~~~~~~~~
  \times e B y \left\{ \frac{2 b (b M_n + y) (2 b M_n + y) (b^2 M_n
  (M^2_n - m^2) + 2 b M^2_n y + y (m + M_n y))}{b^2 (M^2_n - m^2) + 2 b M_n y
  + y^2 + e^{2 (b M_n + y)} (b (m + M_n) + y)^2} \right. \nonumber\\
  &&~~~~~~~~~~~~~~~~~~~~~~~~~~~~~~~~~ \left. + (b^2 M^2_n + 3 b M_n y +
  y^2) \ln \left( 1 + \frac{e^{- 2 (b M_n + y)} (b (- m + M_n) + y)}{b (m +
  M_n) + y} \right) \right\}. 
  \label{PCasgen}
\end{eqnarray}
We plot the behavior of the scaled Casimir pressure in Figs.~\ref{PCas1} and \ref{PCas3}. In general, we can see that its behavior is similar to that of the Casimir energy. From the left panel of Fig.~\ref{PCas1}, one can see the scaled Casimir pressure converges to zero as the increases of parameter $m'$ while from the right panel, it increases as the increases of $\ell^2 |eB|$. These behaviors are supported by Fig.~\ref{PCas3}.  Both panels of Fig.~\ref{PCas1} show that the Casimir pressure increases as the increases of parameter $\lambda$.

We next investigate the Casimir pressure in the case of weak and strong magnetic fields. In the case of weak magnetic field $B\to 0$, the Casimir pressure \eqref{PCasgen} approximately reduces to
\begin{eqnarray}
P_{\rm Cas.}  &\simeq &- \frac{1}{(1 - \lambda) b^4 \pi^2} \int_{b m}^{\infty}
  d x \int_0^{\infty} d v \frac{x^2}{v^{1 / 2} (2 + v)^{3 / 2} }\nonumber\\
  &&
  \times \left( \frac{2 x (1 + v) (2 + v)  (x^2 (1 + v)^2 + t b m - (b
  m)^2)}{x^2 (1 +v)^2 - (b m)^2 + e^{2 x (1 + v)} (b m + x (1 + v))^2}
  + (1 + 3 v + v^2) \ln \left( 1 + \frac{e^{- 2 x (1 + v)} (- b m + x (1 +
  v))}{(b m + x (1 + v))} \right) \right).\nonumber\\
  \label{PB0}
\end{eqnarray}
We further take light mass limit $m'\ll 1$ for the above expression, then we have 
\begin{eqnarray}
P_{\rm Cas.}\simeq  -{(1-\lambda)^2(7\pi^2 (1-\lambda)-80m')\over 960 \ell^4},
\end{eqnarray}
which covers the earlier result of Ref.~\cite{Cruz:2018thz}. As discussed in the previous section, to obtain the above expression, we can use the reverse way, namely, taking its light mass limit and then considering the weak magnetic field.

The Casimir pressure for the case of light mass with the arbitrary magnetic field is approximately given as follows
\begin{eqnarray}
P_{\rm Cas.} \simeq P^{(0)}_{\rm Cas.}+P^{(1)}_{\rm Cas.},
\label{PCasll1}
\end{eqnarray}
where $P^{(0)}_{\rm Cas.}$ is the Casimir pressure for the massless case explicitly given as
\begin{eqnarray}
P^{(0)}_{\rm Cas.}&=& - \sum_{n = 0}^{\infty} i_n \int_0^{\infty} d y \frac{|e B| y}{b^2 \pi^2 (1 -
\lambda) \left( y \left( 2 b \sqrt{2 n e B} + y \right) \right)^{3 / 2}}
\nonumber\\
&&\times \left\{ \frac{2 b \sqrt{2 n e B} \left( 2 b \sqrt{2 n e B} + y \right) \left(
b \sqrt{2 n e B} + y \right) }{\left( 1 + e^{2 \left( b \sqrt{2 n e B} + y
\right)} \right)} + \left( b^2 2 n e B + 3 b \sqrt{2 n e B} y + y^2 \right) \ln \left( 1
+ e^{- 2 \left( b \sqrt{2 n e B} + y \right)} \right) \right\},\nonumber\\
\end{eqnarray}
and $P^{(1)}_{\rm Cas.}$ is the first order correction  to the Casimir pressure $\mathcal{O}(m^\prime)$ explicitly given as
\begin{eqnarray}
 P^{(1)}_{\rm Cas.} =  \sum_{n = 0}^{\infty} i_n \int_0^{\infty} d y \frac{2 |e B| y b \sqrt{2 n e B}
\left( 1 + e^{2 \left( b \sqrt{2 n e B} + y \right)} (1 + 2 y) + 4 e^{2 \left(
b \sqrt{2 n e B} + y \right)} b \sqrt{2 n e B} \right) b m}{b^2 \pi^2 \left( 1
+ e^{2 \left( b \sqrt{2 n e B} + y \right)} \right)^2 \left( y \left( y + 2 b
\sqrt{2 n e B} \right) \right)^{3 / 2} (1 - \lambda) }.
\end{eqnarray}

We next investigate the Casimir pressure \eqref{PCasgen} in the case of heavy mass $m'\gg 1$. In this case,  we have
\begin{eqnarray}
   P_{\rm Cas.}\simeq  - \frac{|e B| \sqrt{m'}}{(1 - \lambda)^{1 / 2} 8 \pi^{3 / 2} b^2} \sum_{n =
0}^{\infty} i_n e^{- 2 \sqrt{{m'}^2 + 2 n e B}},
\label{PCasmgg1}
\end{eqnarray}
and with the limit of the weak magnetic field $B\to 0$,  the above Casimir pressure approximately  reduces to 
\begin{eqnarray}
   P_{\rm Cas.}\simeq  - \frac{ (1 - \lambda)^{5 / 2} {m'}^{3 / 2}}{16 \pi^{3 / 2} \ell^4} e^{-
\frac{{2 m'} }{(1 - \lambda)} }.
\end{eqnarray}
Similar behavior to the Casimir energy \eqref{ECasgg1}, one can see that the Casimir pressure in the limit of heavy mass \eqref{PCasmgg1} converges to zero as increasing of the particle's mass. 

Based on the result of the Casimir pressure in the cases of light \eqref{PCasll1} and heavy masses \eqref{PCasmgg1}, we will analyze the behavior in the strong magnetic field. Taking the limit of strong magnetic field $\ell^2 |eB|\gg 1$ for Eq.~\eqref{PCasll1}, the Casimir pressure approximately reduces to 
\begin{eqnarray}
P_{\rm Cas.}\simeq -{|eB|L^2 (1-\lambda) \over 48 \ell^2},
\label{StrongMmll1}
\end{eqnarray}
while for Eq.~\eqref{PCasmgg1}, we obtain
\begin{eqnarray}
P_{\rm Cas.} \simeq -{|eB|L^2 (1-\lambda)^{3/2} \sqrt{m'} \over 16 \pi^{3/2}  \ell^2 }e^{{-2m'\over (1-\lambda)}}.
\label{StrongMmll1}
\end{eqnarray}
One can also derive both above equations by taking the derivative of the Casimir energy Eqs.~\eqref{ECasll1} and \eqref{ECasgg1} with respect to the plate's distance.

%===============================
\section{Summary}
\label{Summary}
%=======================================================================
We have studied the Casimir effect of a Lorentz-violating Dirac with a background uniform magnetic field. The Lorentz violation is described by two parameters: (i) $\lambda$ , which determines the intensity of the violation and (ii) vector $u^\mu$, which determines the direction of the violation. In the present study, we investigated two vector cases, namely, timelike and spacelike vector cases. For the spacelike vector case, we only discussed the $z$-direction. The purpose of the study is to find the effect of the Lorentz violation parameter $\lambda$ together with the presence of the magnetic field in the behavior of the Casimir energy as well as its pressure. We used the boundary condition from the MIT bag model \cite{Chodos:1974je, Chodos:1974pn, Johnson:1975zp} to represent the property of the plates. From our derivation, we find that for the timelike vector case, the magnetic field and the Lorentz violating parameter do not affect the structure of the momentum constraint while for the spacelike vector case, only Lorentz violating parameter appears.

We noted that the vacuum energy under the MIT boundary condition is divergent.  Using Abel-Plana like summation \cite{Romeo:2000wt}, we can extract this vacuum energy into three main parts, namely, vacuum energy in the absence of the boundary condition, the vacuum energy in the present of single boundary condition that does not relevant to the Casimir effect, and the rest term that refers to the Casimir energy. We can derive the Casimir energy by subtracting the vacuum energy in the presence of the boundary condition from that in the absence of one. The Lorentz violation for the timelike vector case does not affect the structure of the Casimir energy as well as its pressure while for the spacelike vector case, the violation affects it. We also found that the magnetic field has an effect on the Casimir energy and the pressure for both timelike and spacelike vector cases. We have demonstrated the behavior of the scaled Casimir energy and the pressure as a function of mass, parameter $\lambda$, and magnetic field. For the fixed parameter $\lambda$ and magnetic field, the scaled Casimir energy and the pressure converge to zero as the increase of mass (see left panel of Figs.~\ref{EnCas1} and \ref{PCas1}). For fixed parameter $\lambda$ and mass, the scaled Casimir energy and the pressure converge to zero as the increasing of the magnetic field (see right panel of Figs.~\ref{EnCas1} and \ref{PCas1}). We also found that the increase of the parameter $\lambda$ leads to the increase of the Casimir energy and the pressure, as has been pointed out by Ref.~\cite{Cruz:2018thz}. 
For future work, it is interesting to discuss the thermal effect in a similar setup to our present work (c.f.,~Ref.~\cite{Erdas:2021xvv} for the scalar field). It is also interesting to study a similar setup under the general boundary, for example, chiral MIT boundary conditions \cite{Lutken:1983hm}.

%=======================================================================
\section*{Acknowledgments}
A. R. was supported by the  National Research and Innovation Agency (BRIN) Indonesia, through the Post-Doctoral Program. 

\begin{appendix}

%=======================================================================

\section{Detail derivation of 
constraint for momentum}
\label{Detailmomentum}
%=======================================================================
In this section, we provide the complementary derivation for the momentum constraint. Applying the boundary condition from the MIT bag model \eqref{MITBC} to the solution of the Dirac equation, we have two equations as follows
\begin{eqnarray}
&&i\sigma^3\chi_2|_{z=0}-\chi_1|_{z=0}=0,\\
&&i\sigma^3\chi_2|_{z=\ell}+\chi_1|_{z=\ell}=0,
\end{eqnarray}
where we have used $n^{(0)}_\mu=(0,0,0,1)$ and $n^{(\ell)}_\mu=(0,0,0,-1)$ at the first $z=0$ and second plates $z=\ell$, respectively.  Then, in a more explicit expression, we have four equations boundary conditions as follows
\begin{eqnarray}
&&i\chi_{21}|_{z=0}-\chi_{11}|_{z=0}=0,
\label{BC1}\\
&&i\chi_{22}|_{z=0}+\chi_{12}|_{z=0}=0,
\label{BC2}\\
&&i\chi_{21}|_{z=\ell}+\chi_{11}|_{z=\ell}=0,
\label{BC3}\\
&&i\chi_{22}|_{z=\ell}-\chi_{12}|_{z=\ell}=0,
\label{BC4}
\end{eqnarray}
where we have decomposed the two-component spinors $\chi_1$ and $\chi_2$ as 
\begin{eqnarray}
&&\chi_1= \begin{pmatrix}
\chi_{11} \\
\chi_{12}
\end{pmatrix},\\
&&
\chi_2= \begin{pmatrix}
\chi_{21} \\
\chi_{22}
\end{pmatrix}.
\end{eqnarray}
The boundary conditions of Eqs.~\eqref{BC1}-\eqref{BC4} can be simultaneously written in the form of multiplication between two matrices as follows

\begin{eqnarray}
  \begin{pmatrix}
{\cal P}_{11} & {\cal P}_{12} \\
{\cal P}_{21} & {\cal P}_{22}
\end{pmatrix}
  \begin{pmatrix}
  C_0\\
  \tilde C_0
\end{pmatrix}
=0,~~{\rm for}~n=0,
\label{MatrixBoundaryn0}
\end{eqnarray}
and 
\begin{eqnarray}
  \begin{pmatrix}
{\cal Q}_{11} & {\cal Q}_{12} & {\cal Q}_{13} & {\cal Q}_{14}\\
{\cal Q}_{21} & {\cal Q}_{22} & {\cal Q}_{23} & {\cal Q}_{24}\\
{\cal Q}_{31} & {\cal Q}_{32} & {\cal Q}_{33} & {\cal Q}_{34}\\
 {\cal Q}_{41} & {\cal Q}_{42} & {\cal Q}_{43} & {\cal Q}_{44}
\end{pmatrix}
  \begin{pmatrix}
  C_1\\
  C_2\\
  \tilde C_1\\
 \tilde C_2
\end{pmatrix}
=0,~~{\rm for}~n\geq 1,
\label{MatrixBoundary}
\end{eqnarray}
where the matrix elements are given by
\begin{eqnarray}
&&{\cal P}^{(t)}_{11}=ik_3-((1+\lambda)\omega^{(t)}_{0k_3}+m),\\
&&{\cal P}^{(t)}_{12}=-ik_3-((1+\lambda)\omega^{(t)}_{0k_3}+m),\\
&& {\cal P}^{(t)}_{21}=[ik_3+((1+\lambda)\omega^{(t)}_{0k_3}+m)]e^{ik_3\ell},\\
&& {\cal P}^{(t)}_{22}=[-ik_3+((1+\lambda)\omega^{(t)}_{0k_3}+m)]e^{-ik_3\ell},\\
&&{\cal Q}^{(t)}_{11}=-{\cal Q}^{(t)}_{22}=ik_3-((1+\lambda)\omega^{(t)}_{nk_3}+m),\\
&& {\cal Q}^{(t)}_{12}={\cal Q}^{(t)}_{14}={\cal Q}^{(t)}_{21}={\cal Q}^{(t)}_{23}=i\sqrt{2neB},\\
&& {\cal Q}^{(t)}_{13}=-{\cal Q}^{(t)}_{24}=-ik_3-((1+\lambda)\omega^{(t)}_{nk_3}+m),\\
&&{\cal Q}^{(t)}_{31}=-{\cal Q}^{(t)}_{42}=[ik_3+((1+\lambda)\omega^{(t)}_{nk_3}+m)]e^{ik_3\ell},\\
&& {\cal Q}^{(t)}_{32}={\cal Q}^{(t)}_{41}=i\sqrt{2neB}e^{ik_3\ell},\\
&&{\cal Q}^{(t)}_{34}={\cal Q}^{(z)}_{43}=i\sqrt{2neB}e^{-ik_3\ell},\\
&& {\cal Q}^{(t)}_{33}=-{\cal Q}^{(t)}_{44}=[-ik_3+((1+\lambda)\omega^{(t)}_{nk_3}+m)]e^{-ik_3\ell}.
\end{eqnarray}
and 
\begin{eqnarray}
&&{\cal P}^{(z)}_{11}=i(1-\lambda)k_3-(\omega^{(z)}_{0k_3}+m),\\
&&{\cal P}^{(z)}_{12}=-i(1-\lambda)k_3-(\omega^{(z)}_{0k_3}+m),\\
&& {\cal P}^{(z)}_{21}=[i(1-\lambda)k_3+(\omega^{(z)}_{0k_3}+m)]e^{ik_3\ell},\\
&& {\cal P}^{(z)}_{22}=[-i(1-\lambda)k_3+(\omega^{(z)}_{0k_3}+m)]e^{-ik_3\ell},\\
&&{\cal Q}^{(z)}_{11}=-{\cal Q}^{(z)}_{22}=i(1-\lambda)k_3-(\omega^{(z)}_{nk_3}+m),\\
&& {\cal Q}^{(z)}_{12}={\cal Q}^{(z)}_{14}={\cal Q}^{(z)}_{21}={\cal Q}^{(z)}_{23}=i\sqrt{2neB},\\
&& {\cal Q}^{(z)}_{13}=-{\cal Q}^{(z)}_{24}=-i(1-\lambda)k_3-(\omega^{(z)}_{nk_3}+m),\\
&&{\cal Q}^{(z)}_{31}=-{\cal Q}^{(z)}_{42}=[i(1-\lambda)k_3+(\omega^{(z)}_{nk_3}+m)]e^{ik_3\ell},\\
&& {\cal Q}^{(z)}_{32}={\cal Q}^{(z)}_{41}=i\sqrt{2neB}e^{ik_3\ell},\\
&&{\cal Q}^{(z)}_{34}={\cal Q}^{(z)}_{43}=i\sqrt{2neB}e^{-ik_3\ell},\\
&& {\cal Q}^{(z)}_{33}=-{\cal Q}^{(z)}_{44}=[-i(1-\lambda)k_3+(\omega^{(z)}_{nk_3}+m)]e^{-ik_3\ell},
\end{eqnarray}
for timelike and spacelike in the $z$-direction vector cases, respectively.  
For arbitrary non-zero complex coefficients $ C_0,\tilde C_0, C_1, C_2,\tilde C_1,\tilde C_2$ requires the vanishing of the determinant of $2\times 2$ matrix of Eq.~\eqref{MatrixBoundaryn0} and $4\times 4$ matrices of Eq.~\eqref{MatrixBoundary} that lead the constraint for momentum $k_3$.

\section{Negative-energy solutions}
\subsection{Timelike vector case}
\label{NETimelikevectorcase}
The negative energy solution for the right-moving field component is as follows
\begin{eqnarray}
\psi^{(-,t)}_{k_1,n,k_3} ({\bm r})&=& {e^{-ik_1 x}e^{-ik_3 z} \over2\pi\sqrt{2(1+\lambda)\omega^{(t)}_{n, k_3}((1+\lambda) \omega^{(t)}_{n, k_3}+m}) }\nonumber\\
&&\times \left[\tilde C_1
  \begin{pmatrix}
  k_3f^{(t)}_{-k_1 n}(y)\\
  -\sqrt{2neB} f^{(t)}_{-k_1 n-1}(y)\\
  ((1+\lambda)\omega^{(t)}_{nk_3}+m) f^{(t)}_{-k_1 n}(y)\\
  0
\end{pmatrix}
+
\tilde C_2
  \begin{pmatrix}
  -\sqrt{2neB} f^{(t)}_{-k_1 n}(y)\\
  -k_3f^{(t)}_{-k_1 n-1}(y)\\
  0\\
  ((1+\lambda)\omega^{(t)}_{nk_3}+m) f^{(t)}_{-k_1 n-1}(y)
\end{pmatrix}
\right],~~{\rm for}~n\geq 1
\end{eqnarray}
and 
\begin{eqnarray}
\psi^{(-,t)}_{k_1,0,k_3} ({\bm r})= {e^{-ik_1 x}e^{-ik_3 z} \over2\pi\sqrt{2(1+\lambda)\omega^{(t)}_{0, k_3}((1+\lambda) \omega^{(t)}_{0, k_3}+m}) } \tilde C_0 f^{(t)}_{-k_1, 0}(y)
  \begin{pmatrix}
  k_3\\
 0\\
 (1+\lambda)\omega^{(t)}_{0,k_3}+m\\
  0
\end{pmatrix},~~{\rm for}~n=0. 
\end{eqnarray}
The negative energy solution for the left-moving field component is as follows
\begin{eqnarray}
\psi^{(-,t)}_{k_1,n,-k_3} ({\bm r})&=& {e^{-ik_1 x}e^{ik_3 z} \over2\pi\sqrt{2(1+\lambda)\omega^{(t)}_{n, k_3}((1+\lambda) \omega^{(t)}_{n, k_3}+m}) }\nonumber\\
&&\times \left[ C_1
  \begin{pmatrix}
  -k_3f^{(t)}_{-k_1 n}(y)\\
  -\sqrt{2neB} f^{(t)}_{-k_1 n-1}(y)\\
  ((1+\lambda)\omega^{(t)}_{nk_3}+m) f^{(t)}_{-k_1 n}(y)\\
  0
\end{pmatrix}
+
C_2
  \begin{pmatrix}
  -\sqrt{2neB} f^{(t)}_{-k_1 n}(y)\\
  k_3f^{(t)}_{-k_1 n-1}(y)\\
  0\\
  ((1+\lambda)\omega^{(t)}_{nk_3}+m) f^{(t)}_{-k_1 n-1}(y)
\end{pmatrix}
\right],~~{\rm for}~n\geq 1
\end{eqnarray}
and 
\begin{eqnarray}
\psi^{(-,t)}_{k_1,0,-k_3} ({\bm r})= {e^{-ik_1 x}e^{ik_3 z} \over2\pi\sqrt{2(1+\lambda)\omega^{(t)}_{0, k_3}((1+\lambda) \omega^{(t)}_{0, k_3}+m}) } C_0 f^{(t)}_{-k_1, 0}(y)
  \begin{pmatrix}
  -k_3\\
 0\\
 (1+\lambda)\omega^{(t)}_{0,k_3}+m\\
  0
\end{pmatrix},~~{\rm for}~n=0. 
\end{eqnarray}
The total spatial solution inside the confinement area is given by the linear combination between the left- and right-moving field components as follows
\begin{eqnarray}
\psi^{(-,t)}_{k_1,n, l}({\bm r})=\psi^{(-,t)}_{k_1,n,k_{3 l}}({\bm r})+\psi^{(-,t)}_{k_1,n,-k_{3 l}}({\bm r}),
\end{eqnarray}
where we use $k_{3 l}$ to represent the allowed momentum in the system.

\subsection{Spacelike vector case ($z$-direction)}

The negative energy solutions for the right-moving field component  are given as follows
\begin{eqnarray}
\psi^{(-,z)}_{k_1,n,k_3} ({\bm r})= {e^{-ik_1 x}e^{-ik_3 z} \over2\pi\sqrt{2\omega^{(z)}_{n, k_3}(\omega^{(z)}_{n, k_3}+m}) } \left[\tilde C_1
  \begin{pmatrix}
(1-\lambda)  k_3F^{(z)}_{-k_1, n}(y)\\
  -\sqrt{2neB} F^{(z)}_{-k_1, n-1}(y)\\
  (\omega^{(z)}_{n,k_3}+m) F^{(z)}_{-k_1, n}(y)\\
  0
\end{pmatrix}
+
\tilde C_2
  \begin{pmatrix}
  -\sqrt{2neB} F^{(z)}_{-k_1, n}(y)\\
  -(1-\lambda)k_3F^{(z)}_{-k_1, n-1}(y)\\
  0\\
  (\omega^{(z)}_{nk_3}+m) F^{(z)}_{-k_1, n-1}(y)
\end{pmatrix}
\right],~~{\rm for}~n\geq 1\nonumber\\
\end{eqnarray}
and 
\begin{eqnarray}
\psi^{(-,z)}_{k_1,0,k_3} ({\bm r})= {e^{-ik_1 x}e^{-ik_3 z} \over2\pi\sqrt{2\omega^{(z)}_{0, k_3}(\omega^{(z)}_{0, k_3}+m}) }\tilde C_0 F^{(z)}_{-k_1, 0}(y)
  \begin{pmatrix}
  (1-\lambda)k_3\\
 0\\
 \omega^{(z)}_{0, k_3}+m\\
  0
\end{pmatrix}
,~~{\rm for}~n=0, 
\end{eqnarray}
where
\begin{eqnarray}
f^{(t)}_{-k_1, n}(y)=\sqrt{{(eB)^{1/2}\over 2^n n!\pi^{1/2}}}
\exp\Bigg[-{eB \over 2} \bigg(y-{k_1\over eB}\bigg)^2\Bigg]H_n\Bigg[\sqrt{eB}\bigg(y-{k_1\over eB}\bigg)\Bigg]. 
\end{eqnarray}
The negative energy solutions for the left-moving field component  are given as follows
\begin{eqnarray}
\psi^{(-,z)}_{k_1,n,-k_3} ({\bm r})= {e^{-ik_1 x}e^{ik_3 z} \over2\pi\sqrt{2\omega^{(z)}_{n, k_3}(\omega^{(z)}_{n, k_3}+m}) } \left[ C_1
  \begin{pmatrix}
-(1-\lambda)  k_3F^{(z)}_{-k_1, n}(y)\\
  -\sqrt{2neB} F^{(z)}_{-k_1, n-1}(y)\\
  (\omega^{(z)}_{n,k_3}+m) F^{(z)}_{-k_1, n}(y)\\
  0
\end{pmatrix}
+
 C_2
  \begin{pmatrix}
  -\sqrt{2neB} F^{(z)}_{-k_1, n}(y)\\
  (1-\lambda)k_3F^{(z)}_{-k_1, n-1}(y)\\
  0\\
  (\omega^{(z)}_{nk_3}+m) F^{(z)}_{-k_1, n-1}(y)
\end{pmatrix}
\right],~~{\rm for}~n\geq 1\nonumber\\
\end{eqnarray}
and 
\begin{eqnarray}
\psi^{(-,z)}_{k_1,0,-k_3} ({\bm r})= {e^{-ik_1 x}e^{-ik_3 z} \over2\pi\sqrt{2\omega^{(z)}_{0, k_3}(\omega^{(z)}_{0, k_3}+m}) } C_0 F^{(z)}_{-k_1 0}(y)
  \begin{pmatrix}
  -(1-\lambda)k_3\\
 0\\
 \omega^{(z)}_{0, k_3}+m\\
  0
\end{pmatrix}
,~~{\rm for}~n=0. 
\end{eqnarray}

The total spatial solution inside the confinement area is given by the linear combination between the left- and right-moving field components as follows
\begin{eqnarray}
\psi^{(-,z)}_{k_1,n, l}({\bm r})=\psi^{(-,z)}_{k_1,n,k_{3 l}}({\bm r})+\psi^{(-,z)}_{k_1,n,-k_{3 l}}({\bm r}),
\end{eqnarray}
where we use $k_{3 l}$ to represent the allowed momentum in the system.

\end{appendix}

%=======================================================================

\end{document}